\documentclass[11pt,a4paper]{article}
\usepackage{jcappub}
\newcommand{\be}{\begin{equation}}
\newcommand{\ee}{\end{equation}}
\newcommand{\beq}{\begin{eqnarray}}
\newcommand{\eeq}{\end{eqnarray}}

\title{On the Null Trajectories in Conformal Weyl Gravity}

\author[a]{ J. R. Villanueva,\note{Corresponding author.} }
\author[b]{ Marco Olivares }
 \affiliation[a]{\it Departamento de F\'{\i}sica y Astronom\'{\i}a, Universidad de
Valpara\'{\i}so, Gran Breta\~na  1111, Playa Ancha, Valpara\'{\i}so, Chile,}
\affiliation[a] {Centro de Astrof\'isica de Valpara\'iso, Gran Breta\~na  1111, Playa Ancha, Valpara\'{\i}so, Chile.}
\affiliation[b]{\it Instituto de F\'{\i}sica, Pontificia Universidad
Cat\'{o}lica de Valpara\'{\i}so,\\ Av. Universidad 330, Curauma,
Valpara\'{\i}so , Chile.}

 \emailAdd{jose.villanuevalob@uv.cl}
 \emailAdd{marco.olivaresrubilar@gmail.com}

\date{\today}

\abstract{In this work we find analytical solutions to the null geodesics around a black hole in the
conformal Weyl gravity. Exact expressions for the horizons are found, and they depend on
the cosmological constant and the coupling constants of the conformal Weyl gravity. Then,
we study the radial motion from the point of view of the proper and coordinate frames,
and compare it with that found in spacetimes of general relativity. The angular motion
is also examined qualitatively by means of an effective potential; quantitatively,
the equation of motion is solved in terms of $\wp$-Weierstrass elliptic function.
Thus, we find the deflection angle for photons without using any approximation,
which is a novel result for this kind of gravity.}


\begin{document}
\maketitle

\section{Introduction}

For nearly a century, the Einstein's
theory of gravitation \cite{einstein1} has been
studied from various aspects, delivering
many successes both in theoretical
and observational physics, but at the same time, arising many questions.
For example, this theory is not based
on any fundamental principle, it is not
invariant under conformal transformations,
and cannot be described as a quantum field theory.
Besides, the observations of the velocity
distributions in the vicinity of
galaxies is not satisfactory
from the point of view of Einstein's gravitation,
leading to introduction of the dark matter to avoid this problem.
Thus, it makes it desirable to find an alternative
theory of gravity that would repeat the success of Einstein's theory,
but also fix its problems.


A possible candidate is the
Weyl theory, introduced
in 1917 to unify gravity and
electricity \cite{W17}, based on the principle of
local invariance of a manifold, endowed with
the metric $g_{\mu \nu}(x)$, under the change
\begin{equation}
g_{\mu \nu}(x)\rightarrow \Omega^2(x) g_{\mu \nu}(x),
\label{w.0}
\end{equation}
where $\Omega(x)$ is a smooth, strictly positive function.
In a series of works, Mannheim and Kazanas \cite{MK89}
explored the structure of the fourth-order conformal
Weyl gravity (CWG) providing exact solutions to this theory.
Thereafter, much research has been made ​​based on this theory.
For example, the study of black holes solutions can be found
in \cite{klemm98} for topological black holes;
in \cite{said} for static cylindrical black holes; and
\cite{Lu12} for AdS and Lifshitz black holes. Recently,
spherical solutions for charged Weyl black holes has
been found in \cite{Payandeh:2012mj}.
Some cosmological implications of CWG are found in \cite{Knox},
where the authors calculated the abundances of
the primordial light element; also in \cite{Diaferio},
with $\Lambda$CDM model and conformal gravity
put face to face by using the $\gamma$-ray bursts data,
as well as in \cite{Diaferio08} where the authors explain
the properties of X-ray galaxy clusters without
resorting to dark matter. Recently, Mannheim \cite{Mann12}
has studied cosmological perturbations showing the first
steps for the analysis of the tensor fluctuations in CWG.
From the point of view of general relativity test, particularly with
respect to the deflection of light,
much research has been performed. In this sense,
S. Pireaux presents a series of papers giving account
of the critical distances of photons \cite{PI04},
and the constraints on the linear parameters of the theory \cite{PII04}.
After that, based on the works by Rindler and Ishak \cite{RI07,RI10,IRD10},
in which they show that the deflection of light
is influenced by the cosmological constant,
several authors have tried to investigate
the influence of the Weyl parameter in the deflection of light
(see, for example, \cite{Bhattacharya:2009rv,sultana10}).
Unfortunately, this deduction is not entirely exact.
Thus, our interest is to study the allowed motion for massless particles
following the Lagrangian formalism showed in \cite{chandra,COV} to obtain
the exact solution for the trajectories and, consequently, an exact
expression to the bending of light in CWG.

\section{Null Geodesics}
Let us consider a conformal Weyl gravity. An exact static,
spherically symmetric black hole solution is given by \cite{MK89}
\begin{equation}
d\tilde{s}^{2}=-B(\tilde{r})d\tilde{t}^{2}+\frac{d\tilde{r}^{2}}{B(\tilde{r})}+\tilde{r}^{2}(d\theta^{2}+\sin^{2}\theta
d\phi^{2}), \label{w.1}
\end{equation}
where the coordinates are defined in the range $-\infty < \tilde{t} < \infty$,
$\tilde{r}\geq0$, $0\leq\theta\leq\pi$ and $0\leq\phi\leq 2\pi$, and
the lapse function, $B(\tilde{r})$,
is given by
\begin{equation}
B(\tilde{r})=1-\frac{\beta(2-3\beta\,\tilde{\gamma})}{\tilde{r}}-3\beta\,\tilde{\gamma}+\tilde{\gamma} \tilde{r}
- \tilde{k} \tilde{r}^{2}.
\label{w.2}
\end{equation}
Here $\beta$, $\tilde{k}$ and $\tilde{\gamma}$
are positive constants associated to
the central mass, cosmological constant
and the measurements of the departure of
the Weyl theory from the Einstein - de Sitter,
respectively.
Clearly, taking the limit $\tilde{\gamma} =0$
recovers the Schwarzschild - de Sitter (SdS)
case so that we can identify
$\beta=M$ \cite{edery98}.
Is more convenient to work with
dimensionless constants
appearing in the lapse function, by
making the following identifications

\begin{equation}
\frac{d\tilde{s}}{\beta}\rightarrow ds, \quad \frac{\tilde{t}}{\beta}\rightarrow t, \quad  \frac{\tilde{r}}{\beta}\rightarrow r, \quad
\beta \tilde{\gamma} \rightarrow \gamma, \quad \beta^2 \tilde{k} \rightarrow k.\nonumber
\end{equation}
We obtain
\begin{eqnarray}
&&ds^{2}=-B(r)dt^{2}+\frac{dr^{2}}{B(r)}+r^{2}(d\theta^{2}+\sin^{2}\theta\,
d\phi^{2}), \label{w.1.1}\\
&&B(r)=1-\frac{(2-3\gamma)}{r}-3\gamma+\gamma r
- k r^{2},
\label{w.2.1}
\end{eqnarray}
such that the characteristic polynomial of CWG can be written as
\begin{equation}
p_3(r)=-\frac{r B(r)}{k}=r^{3}-\frac{\gamma}{k}r^2 -\frac{(1-3\gamma)}{k}r+\frac{(2-3\gamma)}{k}.
\label{w.3}
\end{equation}
The zeros of the polynomial $p_3(r)$ give us the locations
of the horizons (if there are any). In order to study
the nature of its roots, we perform the following change of variable
\begin{equation}
r=x+\frac{\gamma}{3k}
\label{w.4}
\end{equation}
which yields to
\begin{equation}
p_3(x)=x^3-\eta_2\,x-\eta_3
\label{w.4}
\end{equation}
where the coefficient are given by
\begin{eqnarray}
&&\eta_2=\frac{1}{k}\left(1-3\gamma+\frac{\gamma^2}{3k} \right),\label{w.5.1}\\
&&\eta_3=-\frac{1}{k}\left(1-3\gamma-\frac{\gamma}{3}\left(\frac{1-3\gamma}{k}\right)-\frac{2\gamma^3}{27k^2} \right),
\label{w.5.2}
\end{eqnarray}
and the cubic discriminant is $\Delta_c=27 \eta_3^2-4\eta_2^3$.
Therefore, there are three options for the roots. If
$\Delta_c<0$, there is one real negative root and a
complex pair of the roots, and this represents a naked singularity;
if $\Delta_c>0$, there are three different real roots,
two positive and one negative, which looks similar to the
SdS spacetime with an event and cosmological horizons;
finally, if $\Delta_c=0$, there are three real roots,
a negative one plus a degenerate positive root, and this
represents the extreme case. Assuming that $\gamma$ is small,
one could check that the coefficient $\eta_3$ is negative,
and therefore the discriminant of the polynomial
is positive, $\Delta_c>0$.
Denoting
\begin{equation}
r_w=\frac{\gamma}{3k},\quad R=\sqrt{\frac{\eta_2}{3}},\quad
\varphi=\frac{1}{3}\arccos\left(\frac{|\eta_3|}{R^3}\right),
\end{equation}
we can find the expression for the event horizon, $r_+$,
the cosmological horizon, $r_{++}$, and the negative root
(without physical meaning), $r_n$,
and they have the form
\begin{eqnarray}
&&r_+= r_w+\frac{R}{2}(\cos \varphi-\sqrt{3}\sin\varphi),\label{w.6}\\
&&r_{++}= r_w+R\cos\varphi,\label{w.7}\\
&&r_n= r_w-\frac{R}{2}(\cos \varphi+\sqrt{3}\sin\varphi).\label{w.8}
\end{eqnarray}
Again, the SdS case \cite{jaklitsch} is recuperated for  zero
Weyl parameter ($\gamma=0$).

The motion of photons in this geometry can be determined using
the standard Lagrangian procedure
for geodesic \cite{COV}. In fact, the Lagrangian
for the metric (\ref{w.1}) is given by
\begin{equation}
2\mathcal{L}=-B(r)\dot{t}^{2}+\frac{\dot{r}^{2}}{B(r)}+r^{2}(\dot{\theta}%
^{2}+\sin^{2}\theta \dot\phi^2)\equiv 0,
\label{w.9}
\end{equation}
where the dot denotes a derivative with respect to
the affine parameter $\tau$ along the geodesic.
The equations of motion are obtained from
\be \dot{\Pi}_{q} - \frac{\partial \mathcal{L}}{\partial q} = 0,
\label{w.10} \ee
where $\Pi_{q} = \partial \mathcal{L}/\partial \dot{q}$
are the conjugate momenta to the coordinate $q$.
Since ($t, \phi$) are cyclic coordinates, its
corresponding conjugate momenta are conserved, therefore
\begin{equation}
\Pi_{t} = -B(r) \dot{t} = - \sqrt{E}, \quad
\textrm{and}\quad \Pi_{\phi}
= r^{2}\sin^{2}\theta \dot{\phi} = L,
\label{w.11}
\end{equation}
where $E$ and $L$ are integration constants dimensionless associated to each of them.
Furthermore, these two constants of motion allows us to define the impact
parameter by the relation $b\equiv\frac{L}{\sqrt{E}}$.
Without lack of generality we consider that the motion is developed in the invariant plane
 $\theta = \pi/2$, in which case our set of differential equations are
\begin{eqnarray}
&&\left(\frac{dr}{d\tau}\right)^{2}= E\left(1-\frac{V(r)}{E}\right),\\
\label{w.12}
&&\left(\frac{dr}{d t}\right)^{2}= B^2(r)\left(1-\frac{V(r)}{E}\right),\\
\label{w.13}
&&\left(\frac{dr}{d\phi}\right)^{2}= \frac{r^4}{b^2}\left(1-\frac{V(r)}{E}\right),
\label{w.14}
\end{eqnarray}
where $V(r)$ corresponds to the conformal effective potential defined by
\begin{equation}
V(r)=B(r)\frac{b^{2}E}{r^{2}}=\left(1-\frac{2-3\gamma}{r}-3\gamma+\gamma r
- k r^{2}\right)\frac{b^{2}E}{r^{2}}.
\label{w.16}
\end{equation}

\begin{figure}[tbp]
 \centering
   \includegraphics[width=80mm]{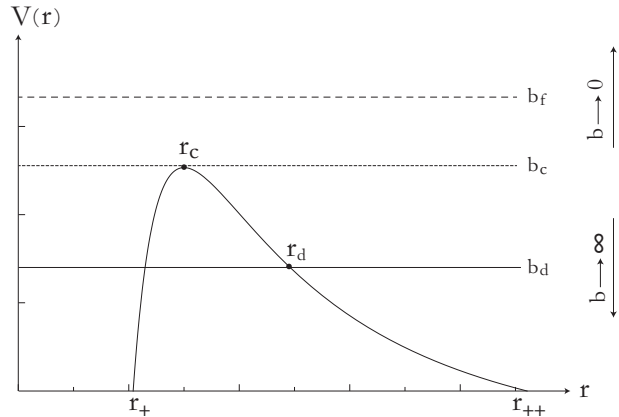}
 \caption{\label{f1} This plot shows the conformal effective potential
  for photons in Weyl's gravity with the values
  of parameters $k=\gamma=0.01$, and $L=3$.
  Since $\gamma$ is small, the spacetime
  looks similar to the Schwarzschild-de Sitter. Note that
  $r_d \rightarrow r_{++}$ when $b \rightarrow \infty$.}
 \end{figure}

A typical plot of the effective potential
(\ref{w.16}) is shown in FIG.\ref{f1}.

\subsection{Radial Motion}

\begin{figure}[!h]
 \begin{center}
   \includegraphics[width=80mm]{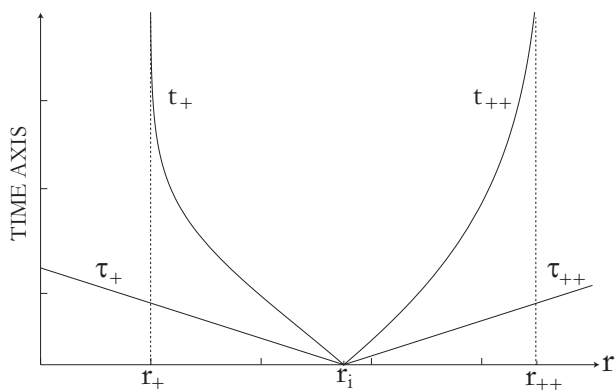}
 \end{center}
 \caption{Plot for the radial motion of photons in CWG. As we can see,
  in the proper frame, the photons arrive to the horizon in a finite time, $\tau_+$ (or $\tau_{++}$) given by
  (\ref{mr.3}) (or (\ref{mr.4})), while
  in the coordinate frame they arrives in an infinite time. This behaviour
  is common with the motion of photons in static spherical symmetric spacetimes of general relativity.}
 \label{f2}
\end{figure}

Radial motion corresponds to a trajectory with null angular
momentum (or zero impact parameter), and photons are destined
to fall toward the event horizon or to the cosmological horizon. From Eq.
(\ref{w.16}) we can see that for radial photons we have
$V(r)=0$, so that eqs. (\ref{w.12}) and (\ref{w.13}) become
\begin{equation}
\frac{dr}{d\tau}=\pm \sqrt{E},
\label{mr.1}
\end{equation}
and
\begin{equation}
\frac{dr}{dt}=\pm B(r),
\label{mr.2}
\end{equation}
respectively, and the sign $+$ ($-$) corresponds to photons  falling
into the cosmological (event) horizon.
Choosing the initial conditions for the photons as $r=r_i$
when $t=\tau=0$, eq. (\ref{mr.1}) yields
\begin{equation}
\tau_+=\frac{r_i-r_+}{\sqrt{E}},
\label{mr.3}
\end{equation}
\noindent and
\begin{equation}
\tau_{++}=\frac{r_{++}-r_i}{\sqrt{E}},
\label{mr.4}
\end{equation}
\noindent which tell us that, in the proper frame of the photons, they
arrive to the event (cosmological) horizon in a finite proper time.
On the other hand, a straightforward integration of eq. (\ref{mr.2}) leads to
\begin{equation}
t(r)=\pm \frac{1}{\sqrt{k}} \ln \left[\left|\frac{r-r_{+}}{r_{i}-r_{+}}\right|^{\mu_1}
\,\left|\frac{r_{++}-r}{r_{+}-r_{i}}\right|^{-\mu_2}\,
\left|\frac{r-r_{n}}{r_{i}-r_{n}}\right|^{\mu_3}\right],
\label{mr.5}
\end{equation}
where the constants are given by
\begin{eqnarray}
&&\mu_1=\frac{r_{+}}{\sqrt{k}(r_{++}-r_{+})(r_{+}-r_n)},\nonumber\\
&&\mu_2=\frac{r_{++}}{\sqrt{k}(r_{++}-r_{+})(r_{++}-r_n)},\nonumber\\
&&\mu_3=-\frac{r_{n}}{\sqrt{k}(r_{++}-r_{n})(r_{+}-r_n)}.\nonumber
\end{eqnarray}
Furthermore, taking the limit $r\rightarrow r_+$  ($r\rightarrow r_{++}$)
in eq. (\ref{mr.5}), it can be shown
that the massless particles take an infinite
coordinate time to cross the event (cosmological) horizon.
This facts result to be common with the spherically
symmetric spacetimes of general relativity \cite{chandra,COV}.
In FIG.\ref{f2}, eqs. (\ref{mr.3}, \ref{mr.4}, \ref{mr.5}) are
plotted.

\subsection{Angular Motion}
When the angular momentum is non-vanishes,  $b\neq 0$, then
$V(r)\neq 0$ in eq. (\ref{w.16}).
A first observation from this conformal effective potential
is that it reaches the maximum
at $r_c=3$ (independently on $\gamma$ and $k$),
which is also  characteristic for the Schwarzschild's spacetimes.
For sure, the impact parameter has a different value at this distance,
giving $b_c^2=27(1+3\gamma-27k)^{-1}$,
which includes the terms from the cosmological constant, $k$, and CWG, $\gamma$.

Next, based on the impact parameter values ​​shown in FIG.\ref{f1},
we present a brief qualitative description of the allowed
angular motions for photons in CWG.
\begin{itemize}
  \item \emph{Capture Zone}:
  If $0<b\equiv b_f <b_{c}$, photons fall inexorably
  to one of the two horizons, depending on initial conditions,
  and its cross section,
  $\sigma$, in these geometry is \cite{wald}
  \begin{equation}\label{mr51}
    \sigma=\pi\,b_c^2=\frac{27\pi}{1+3\gamma-27k}.
  \end{equation}
  \item \emph{Critical Trajectories}:
  If $b=b_{c}$, photons can stay in one of the unstable
  inner circular orbit of radius  $r_{c}=3$.
  Therefore, the photons that arrive from the initial distance
  $r_i$ ($r_+ < r_i< r_c$, or $r_c< r_i<r_{++}$)
  can asymptotically fall to a circle of radius $r_{c}$.
  The proper period in such orbit is
  \begin{equation}\label{p1}
  T_{\tau}=\frac{18\pi}{L},
  \end{equation}
  which results to be the same as the one in the Schwarzschild case \cite{shutz},
  whereas the coordinate period depends on $k$ and $\gamma$ as
  \begin{equation}\label{p2}
  T_t=\frac{6\sqrt{3}\pi}{\sqrt{1+3\gamma-27k}}.
  \end{equation}
  \item \emph{Deflection Zone}. If $b_{c} <b=b_d <\infty $, the
  photons come from a finite  distance $r_i$ ($r_+ < r_i< r_c$ or $r_c< r_i<r_{++}$)
  to a distance $r=r_{d}$ (which is solution of the equation $V(r_d)=E$),
  then return to one of the two horizons. This photons are deflected.
  Also, is possible to find the deflection distance,
  $r_{d}$,  which results to be
  \begin{equation}\label{mr52}
    r_d(b; \gamma, k) = \sqrt{\frac{\varrho_2}{3}}\cos\left[\frac{1}{3}
    \arccos\left(3 \varrho_3\sqrt{\frac{3}{\varrho_2^3}}\right)\right]+\frac{a}{3},
  \end{equation}
  where,  $a=\gamma\, \mathcal{D}^2$, and $\mathcal{D}$ is the {\it anomalous impact parameter}
  given by
  \begin{equation}\label{mr53}
    \mathcal{D}=\frac{b}{\sqrt{1+k\,b^2}}.
  \end{equation}
  We also have
  \begin{equation}\label{mr54}
    \varrho_2=\frac{4\,\mathcal{D}^2}{3}(3-9\gamma+\gamma^2\,\mathcal{D}^2),
  \end{equation}
  and
  \begin{equation}\label{mr55}
    \varrho_3=\frac{4\,\mathcal{D}^2}{27}(-54-27\gamma^2\,\mathcal{D}^2
    +2\gamma^3\,\mathcal{D}^4+9\gamma(9+\mathcal{D}^2)).
  \end{equation}

\end{itemize}
Note that, making $b\rightarrow \infty$ in (\ref{mr53}), drives to
$\varrho_2 = \eta_2$ (c. f. eqs. (\ref{mr54}), (\ref{w.5.1})), and
$\varrho_3 = \eta_3$ (c. f. eqs. (\ref{mr55}), (\ref{w.5.2})), so
we obtain the identity $r_d(\infty; \gamma, k)= r_{++}$.
\begin{figure}[!h]
 \begin{center}
   \includegraphics[width=75mm]{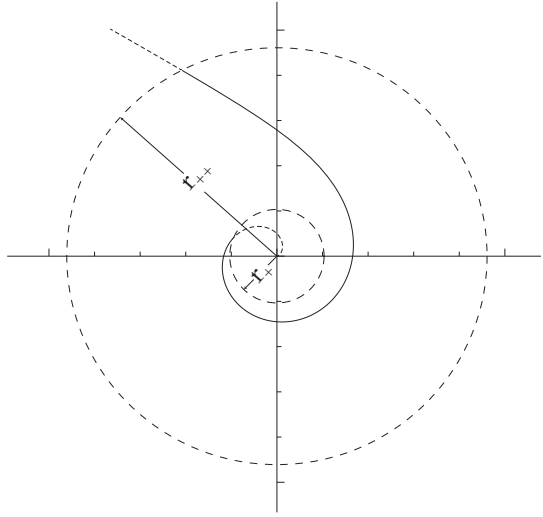}
 \end{center}
 \caption{Polar plot for a capture trajectory of photons.
 Massless particles inevitably fall into one of the horizons}
 \label{f3}
\end{figure}

\begin{figure}[!h]
 \begin{center}
   \includegraphics[width=130mm]{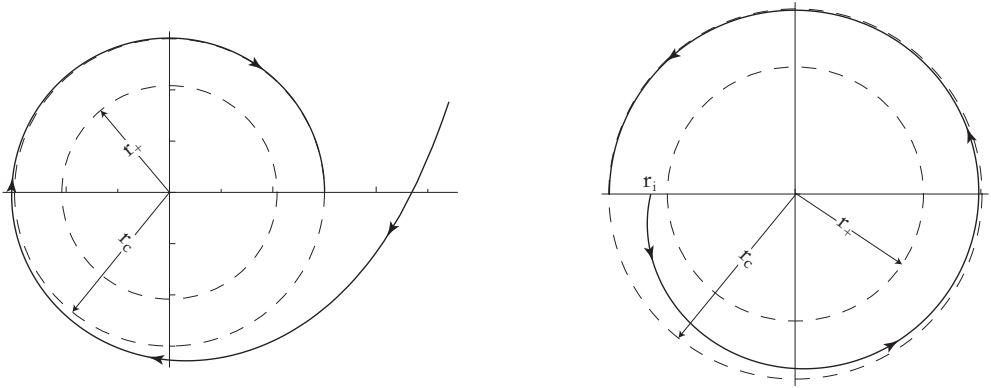}
 \end{center}
 \caption{Polar plot for a critical trajectory of photons coming
  from a distance $r_i$ (LEFT: $r_c<r_i<r_{++}$; RIGHT: $r_+<r_i<r_{c}$).}
 \label{f4}
\end{figure}

With this in mind, after a brief manipulation, it is possible
to integrate out eq. (\ref{w.14}) and obtain
a general solution
\begin{equation}
r(\phi)=\frac{1+\alpha}{4\wp(\phi+\omega_0)+\frac{\alpha}{3}},
\label{mr.6}
\end{equation}
where $\wp(x)\equiv \wp(x; g_2, g_3)$ is the $\wp$-Weierstrass function with
$g_2=1/12$ (again, independent on $k$ and $\gamma$). We also have
in that case
\begin{equation}\label{mr.7}
g_3(b; \gamma, k)= \frac{1}{16}\left[\frac{2}{27}-\gamma^2+\gamma^3+\left(\frac{\alpha}{\mathcal{D}}\right)^2\right],
\end{equation}
and the other parameter is $\alpha=1-3\gamma$, together
with  $\omega_0$ which is an
arbitrary constant of integration.
Note that different orbits are
obtained from (\ref{mr.6}) depending
on the value of the impact parameter
$b$ in (\ref{mr.7}). Thus, the capture
trajectory for $b<b_c$ is showed in FIG.\ref{f3},
while the allowed critical trajectories
for $b=b_c$ are shown in FIG.\ref{f4}.

\section{Bending of light}

\begin{figure}[!h]
 \begin{center}
   \includegraphics[width=130mm]{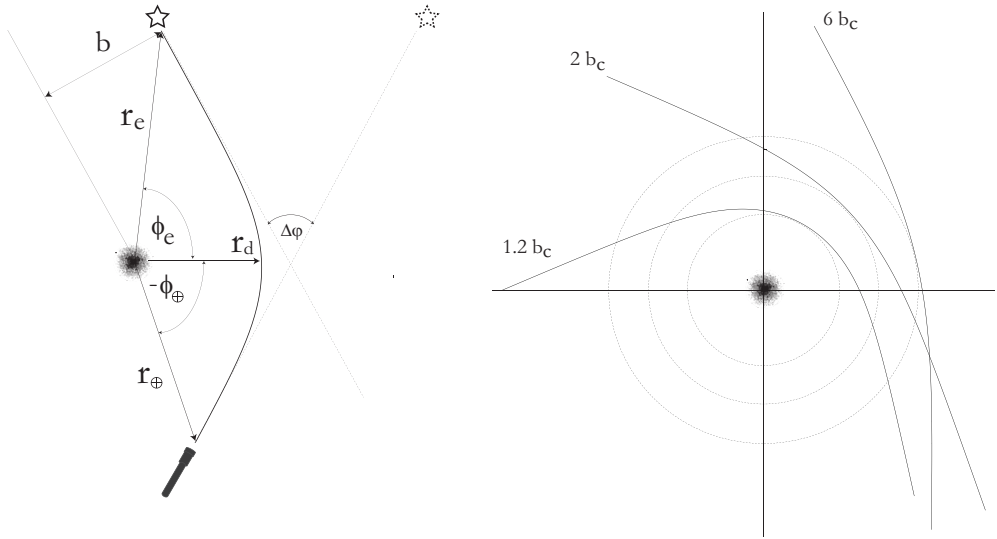}
 \end{center}
 \caption{Polar plot for deflection of light in CWG.
 LEFT: deflection angle, $\Delta \varphi$, for light traveling from a source at $(r_e, \phi_e)$,
 to an observer at $(r_{_{\bigoplus}}, -\phi_{_{\bigoplus}})$;
 RIGHT: graphic for different
 values of the impact parameter, $1.2b_c$, $2b_c$, and $6b_c$, where $b_c=\sqrt{27}(1+3\gamma-27k)^{-1/2}$ is
 the {\it critical impact parameter}.}
 \label{f5}
\end{figure}

On the other hand, from (\ref{mr.6}) and
considering $\Delta\varphi=|\phi_e-(-\phi_{_{\bigoplus}})|-\pi$ (see FIG. \ref{f5}), it is possible to find
the deflection angle accurately, and it reads
\begin{equation}\label{mr.8}
\Delta \varphi =  \left|\wp^{-1}\left[\frac{1+\alpha}{4 r_e}-\frac{\alpha}{12}\right]-\omega_0\right|
+\left|\wp^{-1}\left[\frac{1+\alpha}{4 r_{_{\bigoplus}}}-\frac{\alpha}{12}\right]-\omega_0\right|-\pi,
\end{equation}
where we have taken $r(\phi=0)=r_d$, such that
\begin{equation}\label{mr.9}
  \omega_0=\wp^{-1}\left[\frac{1+\alpha}{4 r_{d}}-\frac{\alpha}{12}\right].
\end{equation}
Here, $\wp^{-1}$ is the inverse $\wp$-Weierstrass function,
$r_e$ is the distance from the central mass
to the source, $r_{_{\bigoplus}}$ is the distance from the central mass
to the earth,
and $r_d$ corresponds to
the deflection distance, see FIG. \ref{f5}. Note that,
unlike what happens in the Schwarzschild spacetime,
there is a maximum distance
determined by the cosmological horizon.
Obviously, the Schwarzschild case
is recovered taking $\gamma=k=0$ in
(\ref{mr.8}). It is also worthwhile noticing
that eq. (\ref{mr.8})
generalizes the result found in
\cite{Bhattacharya:2009rv} derived
via the Rindler-Ishak method \cite{RI07},
in which case one obtains in addition
to the standard deflection angle $ 4M / b $, a
deflection proportional to $ - \gamma b /2 $ ,
which diverges as $ b $ increases.
However, our exact calculation heals this problem,
because when we take the limit $ b \rightarrow \infty $, the deflection distance
tends to the cosmological horizon, which is also possible
to see from the plot of the effective potential, FIG. \ref{f1}, and also from
eq. (\ref{mr52}).
\section{Summary}
In this paper we have found all analytical solutions for null geodesics around a black hole in conformal Weyl gravity.
First, we calculate the horizons of the metric
accurately in terms of the cosmological parameter $k$, and the Weyl parameter, $\gamma$.
Next, we study the radial motion of photons and we determine analytically the behavior of
these while they fall to one of the two horizons, from the point of view of the proper
and coordinate time. We note that photons cross the horizon in a finite proper time,
while in an external system, the photons fall asymptotically to the horizons,
similar to what happens in the Schwarzschild de Sitter black hole.
\newline
We have also studied the angular motion of photons in two ways:
\newline
(i) A qualitative study of the effective potential
(\ref{w.16}) in terms of the impact parameter, $b$,
allows us to find all possible types of
trajectories permitted in this spacetime.
We found that there is a critical value,
$b_c=\sqrt{27}(1+3\gamma-27k)^{-1/2}$, which
corresponds to value of the impact parameter
at $r=r_c=3$, i. e. at the maximum of $V(r)$.
This maximum is common with the Einstein-Kottler spacetimes (S, SdS, SAdS).
Thus, we recognize three families of orbits or zones:
\newline
(a) Capture zone: photons with $b= b_f$, where $0<b_f <b_{c}$,
fall to one of the two horizons with a cross section $\sigma=\pi b_c^2$.
\newline
(b) Critical trajectories: photons with $b=b_c$
can be in an unstable circular orbit of radius $r=r_c$.
Its proper period is the same as that found in the
unstable circular orbit of the Einstein spacetimes,
$T_{\tau}=18\pi/L$. However, we find that
the coordinate period
depends on $k$ and $\gamma$ as
$T_t= T_{t,Sch}/\sqrt{1+3\gamma-27k}$, where
$T_{t,Sch}=6\sqrt{3}\pi$ is the coordinate period
in the unstable circular orbit at the Schwarzschild spacetime.
\newline
(c) Deflection zone: photons with $b=b_d$, where
$b_c<b_d<\infty$, are deflected at a distance $r_d$, which depends
on $k$ and $\gamma$, and also of the impact parameter $b$.
Our exact calculation of $r_d(b; \gamma, k)$ allows us to verify that
$r_d(\infty; \gamma, k)=r_{++}$.
\newline
(ii) The angular motion is integrate ​​in terms of the $\wp$-Weierstrass function,
and thus we obtain the polar form of the trajectories in this geometry, eq. (\ref{mr.6}).
Different trajectories are plotted
according to the impact parameter (see FIG. \ref{f3}, FIG. \ref{f4} and FIG. \ref{f5}).
\newline
Finally, our major goal was to obtain an exact expression for bending of light.
Thus, we obtain the deflection angle in terms of the inverse $\wp$-Weierstrass function,
which results to be a novel result for Conformal Weyl Gravity.

\begin{acknowledgments}
We acknowledge stimulating discussion with Olivera Miskovic (PUCV), V\'ictor
C\'ardenas (UV) and Mario Pedreros (UTA).
M.O. was supported by PUCV through the Proyecto DI Postdoctorado 2012.

\end{acknowledgments}

\end{document}